\documentclass[a4paper]{article}

\usepackage{INTERSPEECH2022}

\usepackage{balance}
\usepackage{xcolor}
\usepackage{multirow}
\usepackage{adjustbox}
\usepackage{soul}
\usepackage{setspace}
\usepackage{eucal}
\usepackage{booktabs}
\usepackage{pifont}
\usepackage{hyperref}

\title{Visually-aware Acoustic Event Detection using Heterogeneous Graphs}
\name{Amir Shirian$^1$, Krishna Somandepalli$^2$, Victor Sanchez$^1$, Tanaya Guha$^3$}
\address{
  $^1$The University of Warwick, UK\\
  $^2$Google Research, USA\\
  $^3$The University of Glasgow, UK}
\email{}

\begin{document}

\maketitle


%
\begin{abstract}
  Perception of auditory events is inherently multimodal relying on both audio and visual cues. A large number of existing multimodal approaches process each modality using modality-specific models and then fuse the embeddings to encode the joint information. In contrast, we employ heterogeneous graphs to explicitly capture the spatial and temporal relationships between the modalities and represent detailed information about the underlying signal. Using heterogeneous graph approaches to address the task of visually-aware acoustic event classification, which serves as a compact, efficient and scalable way to represent data in the form of graphs. Through heterogeneous graphs, we show efficiently modelling of intra- and inter-modality relationships both at spatial and temporal scales. Our model can easily be adapted to different scales of events through relevant hyperparameters. Experiments on \textit{AudioSet}, a large benchmark, shows that our model achieves state-of-the-art performance. Our code is available at \mbox{\href{https://github.com/AmirSh15/VAED_HeterGraph}{\texttt{github.com/AmirSh15/VAED\_HeterGraph}}}
\end{abstract}
\noindent\textbf{Index Terms}: Acoustic event classification, graph neural network, heterogeneous graph, multimodal data. 

\section{Introduction}
\label{sec:intro}
Audio perception by humans is inherently multimodal in nature. It involves processing both aural and visual cues. Visual cues are important not only for audio source localization \cite{atilgan2018integration}, but also for improving audio perception \cite{shi2022robust}. Perceptual studies have also revealed that visual cues can even change how sound is heard \cite{mcgurk1976hearing}.

The majority of existing works on learning audiovisual representations rely on maintaining a tight temporal synchrony between the visual and audio modalities \cite{alwassel2020self,asano2020labelling,korbar2018cooperative}. 
Consider a scene of a bike moving away from the camera. The revving sound of the bike fades as it moves away. While an audio-only-based model may not be capable of detecting the fading sound as 'bike', taking into account the bike as a visual cue, it is possible to identify the event as 'motorbike running'. 
Computer vision-inspired models are common \cite{alayrac2020self,qian2021spatiotemporal,MaZMS21}, where two augmented views of a given audio/audiovisual sample are fed to a shared `backbone', followed by optimizing a contrastive loss \cite{saeed2021contrastive,ma2021contrastive,jiao2020self,jenni2021time,chen2021distilling}, distillation \cite{chen2021distilling,LiuCZLR20}, quantization \cite{alwassel2020self} or information maximization \cite{shukla2020learning,zbontar2021barlow}. However, the vision-inspired audio representation learning methods do not take full advantage of the temporal information available in video data or the complementary knowledge between modalities. Another difficult aspect of such approaches is that data augmentation functions, being vision-inspired, are not often well-suited to a multimodal input.

Heterogeneous graphs are a compact, efficient. and scalable way to represent data involving multiple different entities and their relations \cite{wang2021dualgnn,qian2021dual}. 
Modelling the interaction of entities (including modalities) with heterogeneous graphs is a relatively new paradigm. Multimodal heterogeneous graphs have been successfully used to address various problems in computer vision and natural language processing, such as visual-question answering \cite{saqur2020multimodal}, multimedia recommendation \cite{wei2019mmgcn,wang2021dualgnn}, audio-visual sentiment analysis \cite{yang2021multimodal}, and cross-modal retrieval \cite{qian2021dual}. Multimodal heterogeneous graphs lead to a closer coupling between concepts in multiple modalities, resulting in a significant performance improvement over previous methods \cite{saqur2020multimodal,wei2019mmgcn,yang2021multimodal,wang2021dualgnn}.
Motivated by the success of graph-based methods in multimodal problems
, we propose a heterogeneous graph-based approach to learn visually-aware audio representations. 

\begin{figure*}
    \centering
    \includegraphics[width=1.0\linewidth,height=5cm]{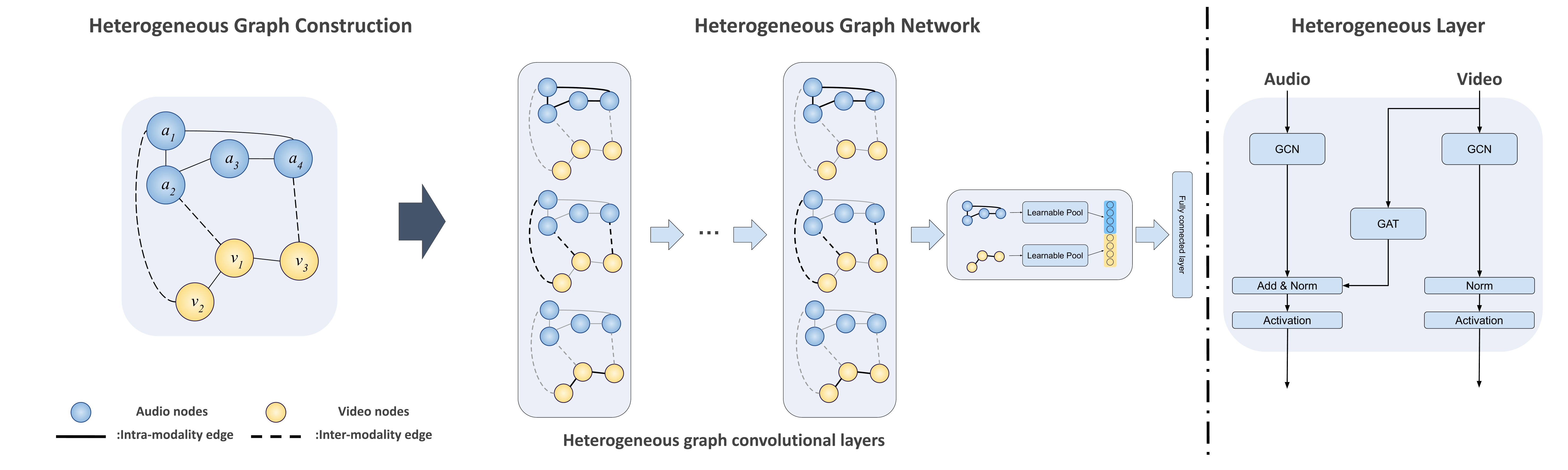}
    \caption{(\textit{Left}) \textbf{Heterogeneous graph architecture}. We split the input audio and video clip into Q and P overlapping segments and then construct the heterogeneous graph containing intra- and inter-modality edges between nodes. Each edge type is considered and processed by the corresponding GNN. For both audio and video modalities, heterogeneous graph convolution layers are utilised to extract the embedding for each node. Separate learnable pooling modules are then used to capture the overall graph representation.
    (\textit{Right}) \textbf{Heterogeneous graph layer} has two independent audio and video flows taking into account intra-modality edges, as well as an attention layer connecting video nodes to audio nodes considering inter-modality edges.
   }
   \vspace{-3mm}
    \label{fig:overview}
\end{figure*}

In this paper, we propose a visually-aware audio representation learning approach based on heterogeneous graphs (see Fig.\ref{fig:overview} for an overview) in the context of acoustic event classification. Our heterogeneous graph model creates a shared space for audio and visual modalities that takes advantage of their spatial and temporal relationships explicitly. We first model the input audiovisual clip as a heterogeneous graph with two sub-graphs, one for each modality with edges capturing inter- and intra-modality relationships. We next develop a heterogeneous graph neural network which is able to capture rich audio representation incorporating complementary information from the visual information. Our contributions are as follows:
\begin{itemize}
    \item We develop a graph construction method for converting an audiovisual clip to a multimodal heterogeneous graph.
    \item We propose a novel heterogeneous graph neural network (HGNN) that can capture modality-specific information as well as complementary information  between modalities.
    \item We demonstrate improved performance by our model for the task of acoustic event classification on the large benchmark AudioSet dataset.
\end{itemize}

\section{Proposed Approach}
\label{sec:proposed}
This section describes our proposed approach for visually-aware audio representation learning. First, we construct heterogeneous graphs to represent the audiovisual data consisting of modality-specific subgraphs and inter-modality edges. Next, we propose a heterogeneous graph neural network (HGNN) architecture that performs graph classification in the context of acoustic event classification.

\subsection{Heterogeneous graph construction}
Our first task is to construct a heterogeneous graph $\mathcal{G} = (\mathcal{V},\mathcal{E},O,R)$, where $\mathcal{V}$ represents the set of nodes, $\mathcal{E}$ the set of edges, $O$ is the set of  node types (object/modality), and $R$ is the set of edge types, where $|O|+|R|>2$. 
Each node $v\in\mathcal{V}$ is associated with a node type and each edge $e\in\mathcal{E}$ is associated with an edge type.
\par
\begin{figure}[t]
    \centering
    \includegraphics[width=1.0\linewidth, trim=0mm 0mm 0mm 0mm, clip=true]{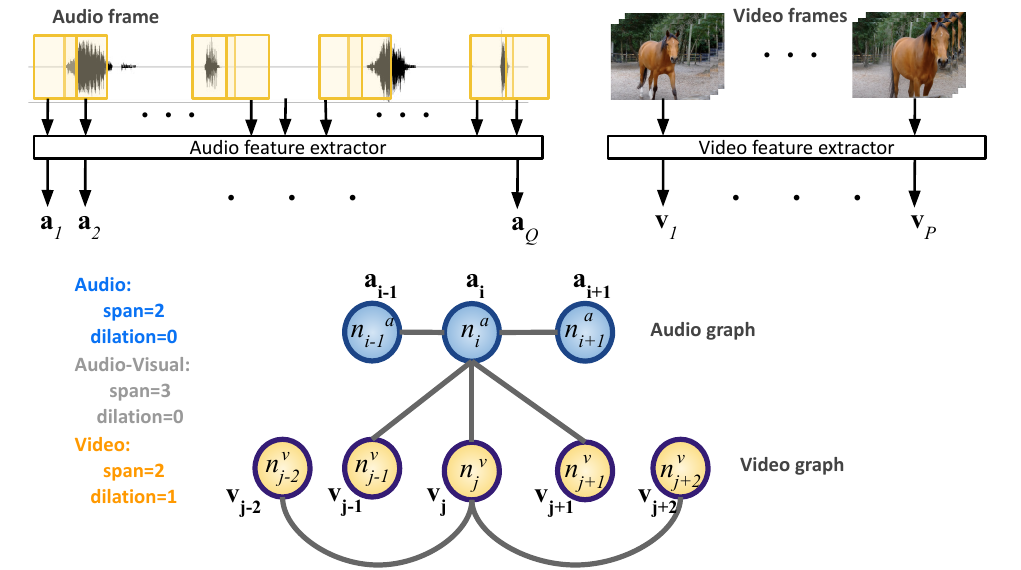}
    \caption{Heterogeneous graph construction process. For simplicity, the edges are only shown for $v_i$ and $v_j$. Similar connections are added for each node.
    }
    \vspace{-5mm}
    \label{fig:graph_const}
\end{figure}

 Given an audiovisual input, we uniformly divide the video frames and the audio into $P$ and $Q$ segments (see  Fig.\ref{fig:graph_const}). The segments are used for feature extraction. Then, given the video %
and the audio segments, %
we  construct a heterogeneous graph with node sets $\mathcal{V}^v = \{v_i\}_{i=1}^P$ and $\mathcal{V}^a = \{a_i\}_{i=1}^Q$, with edge sets $\mathcal{E}=\{\mathcal{E}_{vv},\mathcal{E}_{aa},\mathcal{E}_{va}\}$, which represent edges between video-only nodes, audio-only nodes, and between audio-video nodes respectively. These corresponding adjacency matrices are denoted as $\mathbf{A}_v$, $\mathbf{A}_a$, and $\mathbf{A}_{va}$.  
Each node $v_i\in\mathcal{V}^v$ corresponds to a video segment and its associated feature vector is $\mathbf{n}_i^v\in\mathbb{R}^{d_v}$. Similarly, an audio node $a_i\in\mathcal{V}^a$ is associated with feature vector $\mathbf{n}_i^a\in\mathbb{R}^{d_a}$. Since the graph structure is not naturally defined here, we propose to add inter- and intra-modality edges (see Fig. 2). Additionally, Our graph has two parameters for each edge type, i.e, for $\mathcal{E}_{vv},\mathcal{E}_{aa},\mathcal{E}_{va}$: (i) \textit{span across time} and (ii) \textit{dilation}. The former denotes the number of nodes connected to each node in the temporal direction, whereas the latter denotes leaps between nodes. In total, we have six hyperparameters for graph construction.

\subsection{Heterogeneous graph neural network (HGNN)}
Given heterogeneous graphs $G_1,...,G_N$ and their ground-truth labels $\mathbf{y}_1,...,\mathbf{y}_N$, the task is to learn a $d$-dimensional graph representation $\mathbf{h}_{G_i}\in\mathbb{R}^d$ that captures rich structural and semantic information in $G_i$.

The key idea of most GNNs is to aggregate feature information from a node's neighbours and then update the node feature vector: 
\begin{equation}\label{eq:gnn}
\begin{aligned}
  \mathbf{H}^{k+1} = \sigma\big(\mathbf{A}\mathbf{H}^k\mathbf{W}\big)
\end{aligned}
\end{equation}
where $\mathbf{W}^{(k)}$ is the weight matrix for the $k^{th}$ layer of the GNN, $\sigma$ is a non-linear activation function, such as ReLU, and $k$ is the layer number ($k = 0, \cdots K$). Because of the various node and edge types, this approach is not directly applicable to our heterogeneous graphs.
Previous studies utilise meta-paths for processing heterogeneous graphs \cite{fu2020magnn,hu2020heterogeneous}, which has been shown to be inadequate to properly exploit the information provided by node and edge types \cite{lv2021we}. To overcome this, we use separate GNNs for processing different edge types.

Our HGNN has three flows of information corresponding to the intra- and inter-modality edges as shown in Fig.\ref{fig:overview}. Audio and video flow process the audio and video nodes by considering only intra-modality edges ($\mathcal{E}_{vv},\mathcal{E}_{aa}$) between audio and video nodes, respectively. The third flow carries audio-related information from video nodes to audio nodes for the inter-modality edges ($\mathcal{E}_{va}$):
\begin{equation}\label{eq:flow}
    \begin{aligned}
        \mathbf{n}_{l+1}^a &= \text{GNN}_{\theta_1}\big( \mathbf{n}_l^a, \mathbf{A}_a \big) + \text{GNN}_{\theta_2}\big( \mathbf{n}_l^v, \mathbf{A}_{va} \big)\\
        \mathbf{n}_{l+1}^v &= \text{GNN}_{\theta_3}\big( \mathbf{n}_l^v,\mathbf{A}_v \big)
    \end{aligned}
\end{equation}
where $\mathbf{n}_{l}^a$ and $\mathbf{n}_{l}^v$ are audio and video node features in layer \textit{l}, and GNN is a graph-based neural network such as GCN \cite{kipf2017semi} or GAT \cite{velivckovic2017graph}. The video nodes are only updated using video nodes from the previous layer, as demonstrated in the Eq. \mbox{\ref{eq:flow}}. As audio is the primary source of information in this application, unlike the video branch, the audio nodes are updated using both the audio and video nodes from the preceding layer, bringing information from the video to the audio modality.

Our objective is to classify entire graphs, as opposed to the more common task of classifying each node. Hence, we seek a \emph{graph-level} representation $\mathbf{h}_G\in\mathbb{R}^d$ as the output of our network. This can be obtained by pooling the node-level representations $\mathbf{n}_K^a$, $\mathbf{n}_K^v$ at the $K$-th layer before passing them to the classification layer (see Fig.\ref{fig:overview}). Common choices for pooling functions in the graph domain are \texttt{mean}, \texttt{max}, and \texttt{sum} pooling \cite{kipf2017semi}. Max and mean pooling often fail to preserve the underlying information about the graph structure, while sum pooling has been shown to be a better alternative \cite{xu2018how}. However, all these pooling functions treat adjacent nodes with equal importance, which may not be optimal. To this end and following \cite{shirian2021dynamic}, we propose to \textit{learn} a pooling function $\Psi$ that combines the node embeddings from the $K$-th layer to produce an embedding for the entire graph. The pooling layer for each modality is thus defined as follows: 
\begin{equation} \label{eq:graph_emb}
\begin{aligned}
    \mathbf{h}_G = \Big[\Psi_a(\mathbf{n}_K^a) \,|\,\Psi_v(\mathbf{n}_K^v) \Big]=\mathbf{n}_K^a \mathbf{p}^a + \mathbf{n}_K^v \mathbf{p}^v
\end{aligned}
\end{equation}
where $\mathbf{p}^a$ and $\mathbf{p}^v$ are learnable weights to combine node-level embeddings to obtain a graph-level embedding for audio and video nodes. The overall heterogeneous graph network is trained with focal loss $\mathcal{L}$ as we have a unbalanced dataset:
\begin{equation}\label{eq:loss}
   \mathcal{L} = -\displaystyle\sum_{n} (1-\mathbf{y}_n)^{\gamma} \log \mathbf{\tilde{y}}_n.
\end{equation}

\section{Experiments}
\label{sec:Exp}
\begin{figure*}[tb]
 \centering
 \begin{minipage}[t]{0.33\linewidth}
    \centering
  \includegraphics[width=1\linewidth, trim={0mm 0mm 0mm 0mm}, clip=true, keepaspectratio=false]{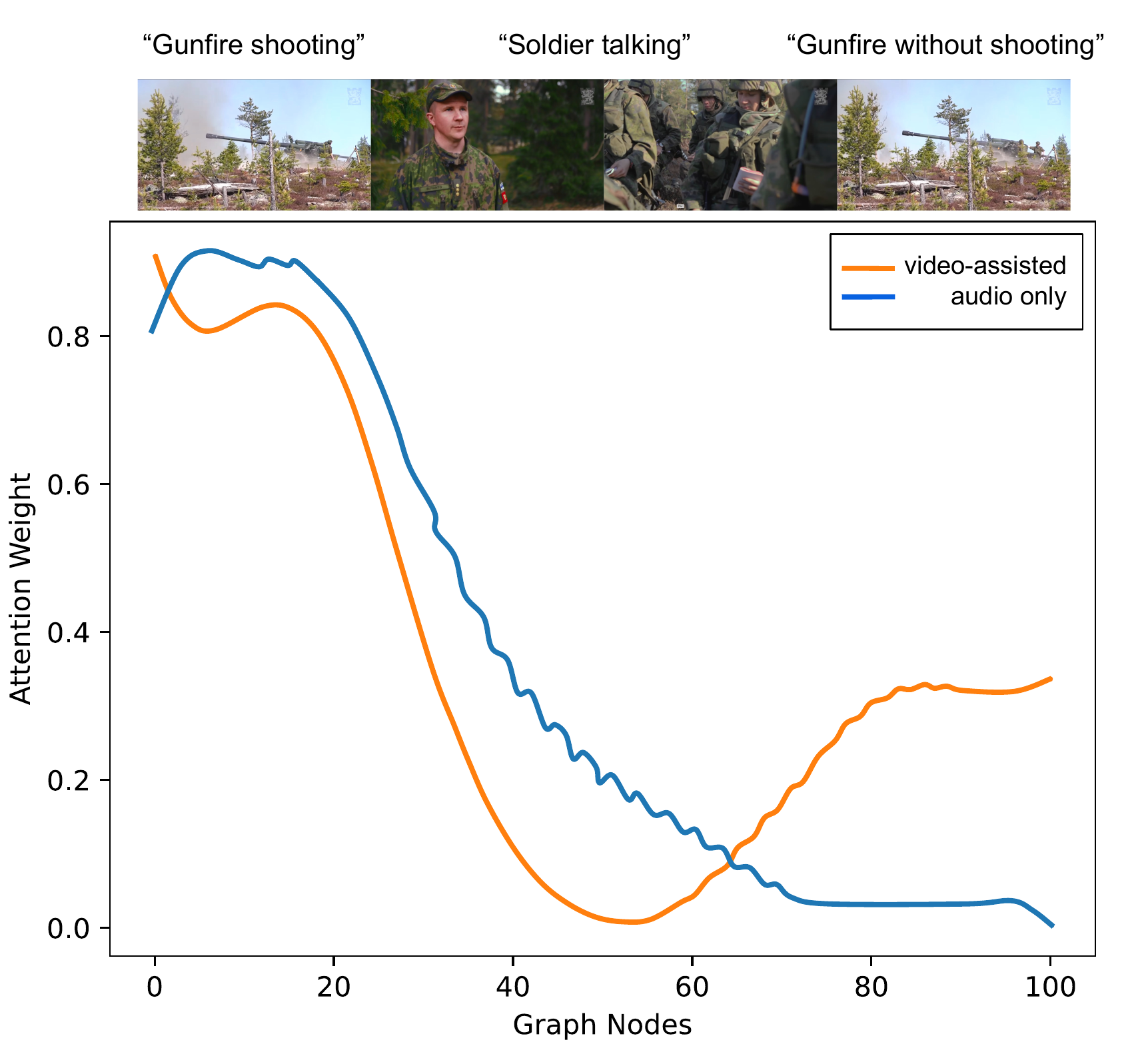}
  
    {\textbf{(a)}} Gunshot, gunfire
 \end{minipage}\hfill
 \begin{minipage}[t]{0.33\linewidth}
    \centering
    \includegraphics[width=1\linewidth, trim={0mm 0mm 0mm 0mm}, clip=true, keepaspectratio=false]{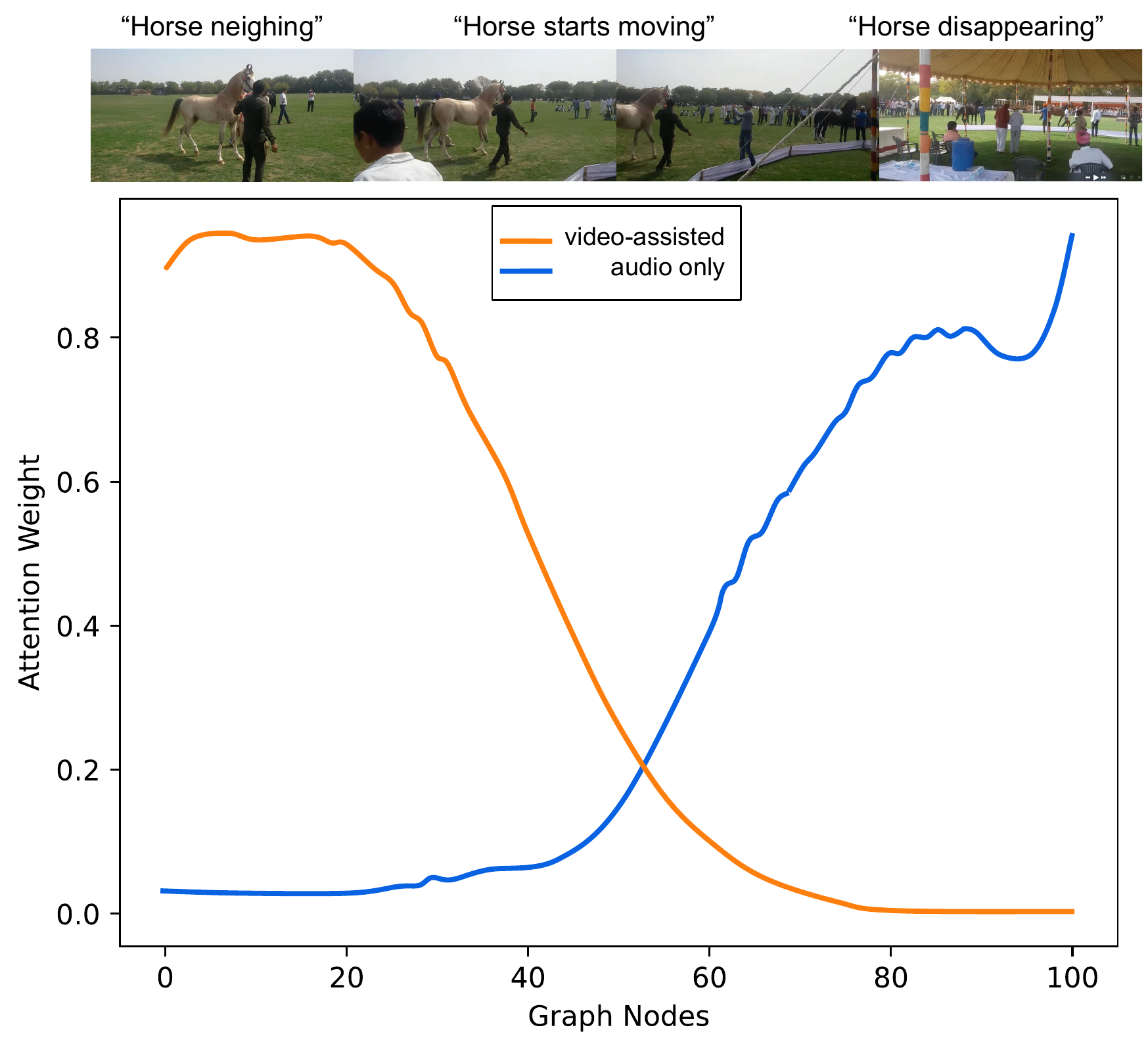}
    
    {\textbf{(b)}} Horse neighing
 \end{minipage}\hfill
 \begin{minipage}[t]{0.33\linewidth}
    \centering
  \includegraphics[width=1\linewidth, trim={0mm 0mm 0mm 0mm}, clip=true, keepaspectratio=false]{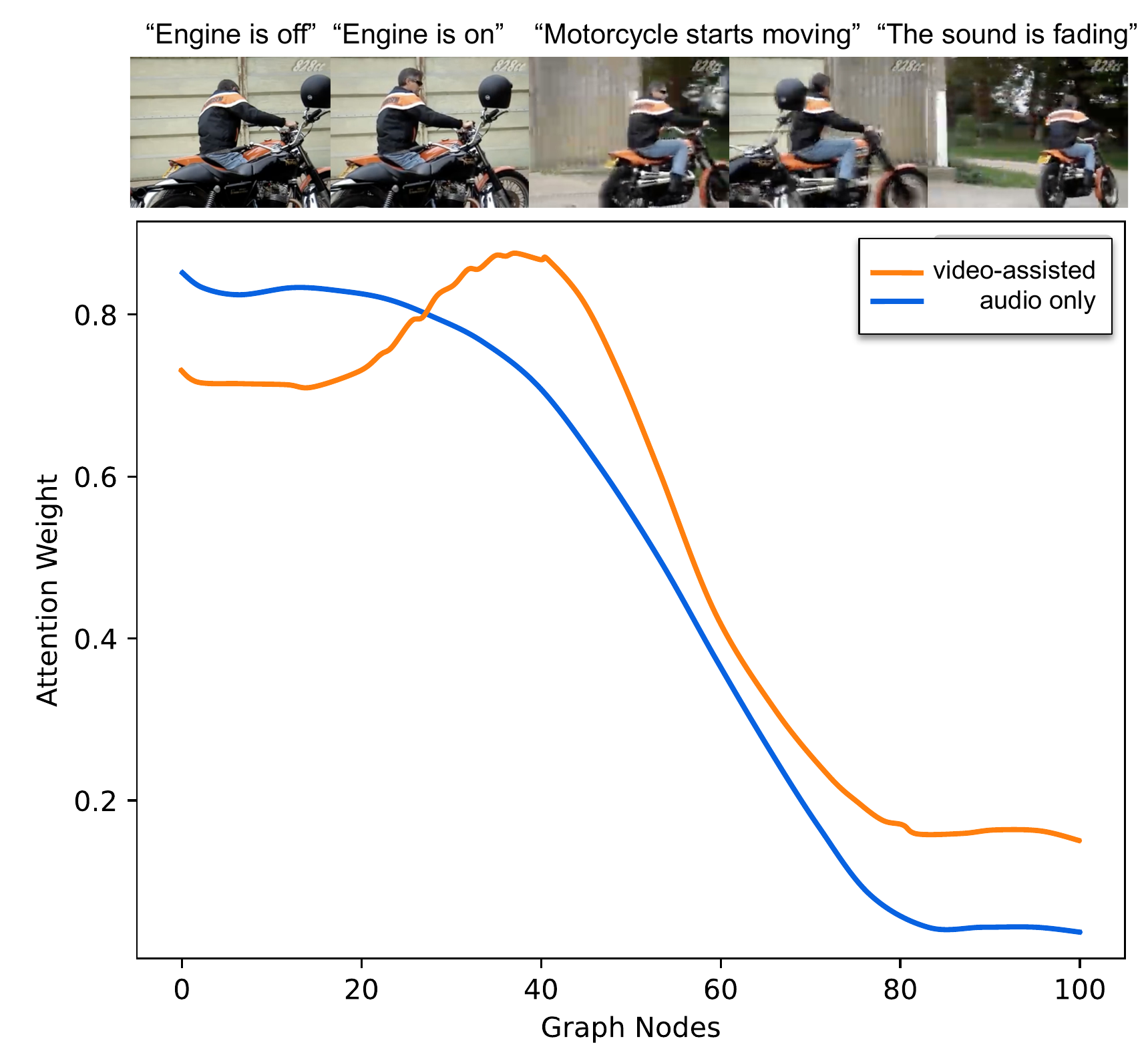}
  
    {\textbf{(c)}} Motorcycle starting
 \end{minipage}\hspace{1mm}
 \caption{
 Qualitative results showing attention weights corresponding to the audio nodes for with (in orange) and without (in blue) video supervision. Each node represents a segment of 100-millisecond duration and the ground-truth label for each video is provided below. Attention values were normalized and re-scaled to [0,1] range.
 (a) This video begins with a strong machine firing sound. After that, a soldier is interrogated, followed by footage of troops. Finally, the machine appears again but is not fired. Even without the associated sound of shooting, the video-assisted audio nodes are able to recognise the firing machine towards the end by assigning higher attention weights to these moments.
 (b) A horse begins neighing and moves away from the camera. As it moves, the sound fades. The horse is no longer visible or audible as time elapses. The audio-only model incorrectly detects these moments by assigning high attention values, while the video-assisted model correctly discards these moments.
 (c) A video of a motorcycle moving. The video-assisted attention weights suggest that our model can capture additional meaningful patterns, such as the engine start. Furthermore, as the engine sound fades, the attention weights corresponding to the audio-only model decrease, and the video-assisted attention weights have relatively higher values indicating that the video information extracted by our model is complementary to the audio event information.}
 \label{fig:qualitative}
 \vspace{-4mm}
\end{figure*}

In this section, we first discuss the dataset used for benchmarking and feature extraction details. We then present experimental results and analysis to evaluate the performance of the proposed HGNN architecture.
\subsection{Dataset}
\vspace{-1mm}
We use a large scale weakly labelled dataset \textbf{AudioSet} \cite{gemmeke2017audio}, which contains audio segments from YouTube videos. We work with 33 categories from the balanced set that have high rater confidence score ($\{0.7, 1.0\}$). This yields a training set of 82,410 clips. For a fair comparison with baseline methods, we also use the original evaluation set, which has 85,487 test clips.

\subsection{Feature Encoder}
\vspace{-1mm}
\noindent \textbf{Audio Encoder.} To extract the audio node features, each audio clip is divided into 960 ms segments with 764 ms overlap. For each segment, a log-mel spectrogram is computed by taking its short-time Fourier transform using a frame of 25 ms with 10ms overlap, 64 mel-spaced frequency bins, and log-transforming the magnitude of each bin. This creates log-mel spectrograms of dimensions $96\times64$, which are the input to the pre-trained VGGish network \cite{hershey2017cnn}. 
We use the 128-dimensional features extracted by the VGGish network for each log-mel spectrogram.

\noindent \textbf{Video Encoder.} Each video is segmented into non-overlapping 250 ms chunks to extract the video node features. The 1024-dimensional feature is then obtained by feeding each segment into an off-the-shelf 3D convolution network, S3D \cite{xie2017rethinking} (trained with self-supervision\cite{han2020self}). Note that our method is not limited to these pre-trained embeddings and can work with any generic embeddings for both audio and video.

\subsection{Implementation Details}
\vspace{-1mm}
Each video clip produces a heterogeneous graph with $P=40$ audio and $Q=100$ video nodes, where each node corresponds to a 960 ms length audio or 250 ms length video segment. 
We repeat our experiments $10$ times with different seeds and report both mAP (mean average precision) and ROC-AUC (area under the ROC curve) values. 
Our network weights are initialized following the Xavier initialization. We used Adam optimizer with a learning rate of $0.005$, a decay rate of $0.1$ after $1500$ iterations, and $1000$ warm-up iterations for all experiments. We set $\gamma=2$ (see Eq. \ref{eq:loss}). The graph construction hyper-parameters are explored heuristically and set to \textit{span audio} = 6, \textit{dilation audio} = 3, \textit{span video} = 4, \textit{dilation video} = 4, \textit{span audio-visual} = 3, and \textit{dilation audio-visual} = 1 for all experiments. For graph neural network, we select regular GCNs \cite{kipf2017semi} for each modality branch and a GAT \cite{velivckovic2017graph} for fusing information from video to audio branch, resulting in $4$ heterogeneous layers (Fig. \ref{fig:overview}) with a hidden size of $512$ for all layers.
We use Pytorch on an NVIDIA RTX-2080Ti GPU. 

\begin{table}[t]
\centering
\caption{Acoustic event classification results on \textbf{AudioSet}}
\resizebox{1.0\linewidth}{!}{
\renewcommand*{\arraystretch}{1.0}
  \begin{tabular}{l|c|c|c}\cline{2-4}
        \toprule
        \bf Model & \bf mAP & \bf ROC-AUC & \bf Params          \\ 
        \midrule
        Ours audio only    & $0.42\pm0.01$ & $0.90\pm0.00$ & $1.4$M\\   
        Ours video only & $0.15\pm0.02$ & $0.75\pm0.01$ & $1.5$M\\
        \textbf{Ours} both & $\mathbf{0.50} \pm0.01$ & $0.93\pm0.00$ & $2.1$M\\
        \midrule
        \multicolumn{4}{c}{\emph{Baselines}}  \\
        \midrule
        ResNet-1D audio only & $0.35 \pm0.01$ & $0.90\pm0.00$ & $40.4$M\\
        ResNet-1D both & $0.38 \pm0.03$ & $0.89 \pm0.02$ & $81.2$M\\
        LSTM audio only & $0.40 \pm0.00$ & $0.90\pm0.00$ & $0.8$M\\
        \midrule
        \multicolumn{4}{c}{\emph{State-of-the-art}}  \\
        \midrule
        DaiNet \cite{dai2017very} & $0.25\pm0.07$ & - & $1.8$M \\
        Spectrogram-VGG   & $0.26\pm0.01$ & - & $6$M\\
        VATT \cite{akbari2021vatt} & $0.39\pm0.02$ & - & $87$M  \\
        SSL graph \cite{shirian2022self}   & $0.42\pm0.02$  & - & $218$K  \\
        Wave-Logmel \cite{kong2020panns} & $0.43\pm0.04$  & - & $81$M  \\
       AST \cite{gong2021ast} & $0.44\pm0.00$ & -  & $88$M  \\
        \bottomrule
        \end{tabular}
        }
        \label{tab:Auco_Eve_Clas}
        \vspace{-3.5mm}
\end{table}

\subsection{Results and analysis}
\noindent \textbf{Baselines.} We compare our method with a number of fully and self-supervised models, as tabulated in Table \ref{tab:Auco_Eve_Clas}.
The Spectrogram-VGG model is the same as configuration A in \cite{simonyan2014very}, with only one change: the final layer is a softmax with 33 units. The feature for each audio input to the VGG model is a log-mel spectrogram of dimensions 96$\times$64 computed by averaging across non-overlapping segments of length 960ms. We also compared our method with a graph-based work. Each node in this work represents an audio clip, and a KNN subgraph has been created, as well as a GNN that is trained using graph self-supervised proxy tasks \cite{shirian2022self}. We also use the two popular spatial and temporal network architectures, ResNet-1D \cite{hong2020holmes} and LSTM, with pretrained embedding features for both audio and video as input, to further investigate the superiority of our graph modelling. All baseline hyper-parameters are set to the values published in the original papers.
Note that we do not utilise any data augmentation, despite the fact that other methods used powerful data augmentations. Additionally, all of the baselines have been retrained using the same classes as our model.

\noindent \textbf{Results.}
Table \ref{tab:Auco_Eve_Clas} reports the mAP and ROC-AUC (averaged over 10 runs with different seeds) values with standard deviation for each model and their variants. It compares the performance of our model with different independent modalities and strong baselines with that of the heterogeneous model in terms of mAP. The heterogeneous graph model outperforms the homogeneous graph and non-graph models. Our method leverages the pre-trained features as node attributes. Thus, to check the performance of our graph-based model, two strong baselines, ResNet-1D and LSTM, have been selected. Compared to these methods, our homogeneous graph sub-models achieve a superior mAP score that demonstrates the effectiveness of our graph-based modelling strategy. Furthermore, when compared to other baselines, our heterogeneous graph-based model achieves the greatest ROC-AUC score ($0.93$), implying more trustworthy predictions at various thresholds.
When compared with the other supervised models, our heterogeneous graph model outperforms Spectrogram-VGG and DaliNet \cite{dai2017very}. Our model also has significantly fewer learnable parameters compared with the recent transformer-based architectures, VATT and AST. 

\noindent \textbf{Ablation experiments.} 
We perform exhaustive ablation experiments to investigate the contribution of each component we propose to build our heterogeneous graph neural network. Table~\ref{tab:ablation} presents the ablation results on the AudioSet dataset. We observe that each new component brings improvement. In all experiments, model performance is measured with mAP to quantify the recognition rate. The introduction of the heterogeneous graph increases the recognition rate by about $9\%$; when combined with our new graph attentional convolution layer between modalities (right half of Fig.~\ref{fig:overview}), the performance increases to $0.49$. Adding the learnable pooling brings up the mAP score to $0.50$. Removing the learnable pooling however reduces the performance by about $3\%$ and $1\%$ for audio-only and video-only models, respectively.
The ablation results show that each of the proposed components in our architecture is important, and contributes positively towards the overall model performance.

\begin{table}[tb]
\centering
\caption{Ablation experiments on the AudioSet dataset. Each new component in our heterogeneous network contributes towards its performance.}
\resizebox{0.8\linewidth}{!}{
\label{tab:ablation}
\vspace{-1mm}
\renewcommand*{\arraystretch}{1.2}
\begin{tabular}{cccc c}
\hline
{\bf Audio} & {\bf Video } & \bf  Attn  & \bf Learned $\mathbf{p}$ & {\bf mAP}  \\ 
\hline \hline
\checkmark & - & -     &  - & $0.38$          \\
\checkmark & - & -     &  \checkmark & $0.41$          \\
- & \checkmark & -     &  - & $0.12$          \\
- & \checkmark & -     &  \checkmark & $0.13$          \\
\checkmark & \checkmark & -  &  - & $0.49$        \\
\checkmark & \checkmark & \checkmark  &  - & $0.49$        \\
 \checkmark & \checkmark & \checkmark    & \checkmark &  $\mathbf{0.50}$\\
\hline
\end{tabular}
}
\vspace{-4mm}
\end{table}

\noindent \textbf{Qualitative results.} 
We display how our model attends to different nodes to gain insights into its learning process. Because each video clip is divided into 100ms segments, each node represents a 100ms time window. In Fig.~\ref{fig:qualitative}, we show the attention weights corresponding to audio nodes in cases of with and without video supervision for three input videos from the test set with distinct acoustic classes. Then, for each video, we sample four frames and display them on top of each figure to provide more visual information. This gives rise to \emph{salient} nodes for each input. The results show that the proposed model can extract visually complementary information to an audio event from heterogeneous graphs as input.

\section{Conclusion}
\label{sec:conclusions}
In this paper, we introduced the idea of hetergeneous graphs to model audio data with visual cues. We proposed a compact and efficient graph-based architecture that learns audio representations effectively in the context of acoustic event classification. We transformed an audiovisual input to a heterogeneous graph with different learnable hyper-parameters capturing intra and inter modalities connections in both spatial and temporal domains. Our heterogeneous graph model produces higher or comparable performance to the state-of-the-art on a popular benchmark dataset, the AudioSet. Our current model relies on pre-trained embeddings, which gives the flexibility of choosing any suitable embeddings. Nevertheless, our model can be made end-to-end trainable, which will be addressed as part of our future work.

\bibliographystyle{IEEEtran}

\bibliography{biblo}

\begin{thebibliography}{10}
\providecommand{\url}[1]{#1}
\csname url@samestyle\endcsname
\providecommand{\newblock}{\relax}
\providecommand{\bibinfo}[2]{#2}
\providecommand{\BIBentrySTDinterwordspacing}{\spaceskip=0pt\relax}
\providecommand{\BIBentryALTinterwordstretchfactor}{4}
\providecommand{\BIBentryALTinterwordspacing}{\spaceskip=\fontdimen2\font plus
\BIBentryALTinterwordstretchfactor\fontdimen3\font minus
  \fontdimen4\font\relax}
\providecommand{\BIBforeignlanguage}[2]{{%
\expandafter\ifx\csname l@#1\endcsname\relax
\typeout{** WARNING: IEEEtran.bst: No hyphenation pattern has been}%
\typeout{** loaded for the language `#1'. Using the pattern for}%
\typeout{** the default language instead.}%
\else
\language=\csname l@#1\endcsname
\fi
#2}}
\providecommand{\BIBdecl}{\relax}
\BIBdecl

\bibitem{atilgan2018integration}
H.~Atilgan, S.~M. Town, K.~C. Wood, G.~P. Jones, R.~K. Maddox, A.~K. Lee, and
  J.~K. Bizley, ``Integration of visual information in auditory cortex promotes
  auditory scene analysis through multisensory binding,'' \emph{Neuron},
  vol.~97, no.~3, pp. 640--655, 2018.

\bibitem{shi2022robust}
B.~Shi, W.-N. Hsu, and A.~Mohamed, ``Robust self-supervised audio-visual speech
  recognition,'' \emph{arXiv preprint arXiv:2201.01763}, 2022.

\bibitem{mcgurk1976hearing}
H.~McGurk and J.~MacDonald, ``Hearing lips and seeing voices,'' \emph{Nature},
  vol. 264, no. 5588, pp. 746--748, 1976.

\bibitem{alwassel2020self}
H.~Alwassel, D.~Mahajan, B.~Korbar, L.~Torresani, B.~Ghanem, and D.~Tran,
  ``Self-supervised learning by cross-modal audio-video clustering,''
  \emph{NeurIPS}, vol.~33, pp. 9758--9770, 2020.

\bibitem{asano2020labelling}
Y.~Asano, M.~Patrick, C.~Rupprecht, and A.~Vedaldi, ``Labelling unlabelled
  videos from scratch with multi-modal self-supervision,'' \emph{NeurIPS},
  vol.~33, pp. 4660--4671, 2020.

\bibitem{korbar2018cooperative}
B.~Korbar, D.~Tran, and L.~Torresani, ``Cooperative learning of audio and video
  models from self-supervised synchronization,'' \emph{NeurIPS}, vol.~31, 2018.

\bibitem{alayrac2020self}
J.-B. Alayrac, A.~Recasens, R.~Schneider, R.~Arandjelovi{\'c}, J.~Ramapuram,
  J.~De~Fauw, L.~Smaira, S.~Dieleman, and A.~Zisserman, ``Self-supervised
  multimodal versatile networks,'' \emph{NeurIPS}, vol.~33, pp. 25--37, 2020.

\bibitem{qian2021spatiotemporal}
R.~Qian, T.~Meng, B.~Gong, M.-H. Yang, H.~Wang, S.~Belongie, and Y.~Cui,
  ``Spatiotemporal contrastive video representation learning,'' in \emph{CVPR},
  2021, pp. 6964--6974.

\bibitem{MaZMS21}
S.~Ma, Z.~Zeng, D.~J. McDuff, and Y.~Song, ``Active contrastive learning of
  audio-visual video representations,'' in \emph{{ICLR}}, 2021.

\bibitem{saeed2021contrastive}
A.~Saeed, D.~Grangier, and N.~Zeghidour, ``Contrastive learning of
  general-purpose audio representations,'' in \emph{ICASSP}.\hskip 1em plus
  0.5em minus 0.4em\relax IEEE, 2021, pp. 3875--3879.

\bibitem{ma2021contrastive}
S.~Ma, Z.~Zeng, D.~McDuff, and Y.~Song, ``Contrastive learning of global and
  local video representations,'' \emph{NeurIPS}, vol.~34, 2021.

\bibitem{jiao2020self}
J.~Jiao, Y.~Cai, M.~Alsharid, L.~Drukker, A.~T. Papageorghiou, and J.~A. Noble,
  ``Self-supervised contrastive video-speech representation learning for
  ultrasound,'' in \emph{International Conference on Medical Image Computing
  and Computer-Assisted Intervention}.\hskip 1em plus 0.5em minus 0.4em\relax
  Springer, 2020, pp. 534--543.

\bibitem{jenni2021time}
S.~Jenni and H.~Jin, ``Time-equivariant contrastive video representation
  learning,'' in \emph{ICCV}, 2021, pp. 9970--9980.

\bibitem{chen2021distilling}
Y.~Chen, Y.~Xian, A.~Koepke, Y.~Shan, and Z.~Akata, ``Distilling audio-visual
  knowledge by compositional contrastive learning,'' in \emph{CVPR}, 2021, pp.
  7016--7025.

\bibitem{LiuCZLR20}
M.~Liu, X.~Chen, Y.~Zhang, Y.~Li, and J.~M. Rehg, ``Attention distillation for
  learning video representations,'' in \emph{31st British Machine Vision
  Conference 2020, {BMVC}}, 2020.

\bibitem{shukla2020learning}
A.~Shukla, S.~Petridis, and M.~Pantic, ``Learning speech representations from
  raw audio by joint audiovisual self-supervision,'' \emph{arXiv preprint
  arXiv:2007.04134}, 2020.

\bibitem{zbontar2021barlow}
J.~Zbontar, L.~Jing, I.~Misra, Y.~LeCun, and S.~Deny, ``Barlow twins:
  Self-supervised learning via redundancy reduction,'' in \emph{ICML}.\hskip
  1em plus 0.5em minus 0.4em\relax PMLR, 2021, pp. 12\,310--12\,320.

\bibitem{wang2021dualgnn}
Q.~Wang, Y.~Wei, J.~Yin, J.~Wu, X.~Song, L.~Nie, and M.~Zhang, ``Dualgnn: Dual
  graph neural network for multimedia recommendation,'' \emph{IEEE Transactions
  on Multimedia}, 2021.

\bibitem{qian2021dual}
S.~Qian, D.~Xue, H.~Zhang, Q.~Fang, and C.~Xu, ``Dual adversarial graph neural
  networks for multi-label cross-modal retrieval,'' in \emph{Thirty-Fifth AAAI
  Conference on Artificial Intelligence, AAAI}, 2021, pp. 2440--2448.

\bibitem{saqur2020multimodal}
R.~Saqur and K.~Narasimhan, ``Multimodal graph networks for compositional
  generalization in visual question answering,'' \emph{NeurIPS}, vol.~33, pp.
  3070--3081, 2020.

\bibitem{wei2019mmgcn}
Y.~Wei, X.~Wang, L.~Nie, X.~He, R.~Hong, and T.-S. Chua, ``Mmgcn: Multi-modal
  graph convolution network for personalized recommendation of micro-video,''
  in \emph{Proceedings of the 27th ACM International Conference on Multimedia},
  2019, pp. 1437--1445.

\bibitem{yang2021multimodal}
X.~Yang, S.~Feng, Y.~Zhang, and D.~Wang, ``Multimodal sentiment detection based
  on multi-channel graph neural networks,'' in \emph{IJCNLP}, 2021, pp.
  328--339.

\bibitem{fu2020magnn}
X.~Fu, J.~Zhang, Z.~Meng, and I.~King, ``Magnn: Metapath aggregated graph
  neural network for heterogeneous graph embedding,'' in \emph{Proceedings of
  The Web Conference 2020}, 2020, pp. 2331--2341.

\bibitem{hu2020heterogeneous}
Z.~Hu, Y.~Dong, K.~Wang, and Y.~Sun, ``Heterogeneous graph transformer,'' in
  \emph{Proceedings of The Web Conference 2020}, 2020, pp. 2704--2710.

\bibitem{lv2021we}
Q.~Lv, M.~Ding, Q.~Liu, Y.~Chen, W.~Feng, S.~He, C.~Zhou, J.~Jiang, Y.~Dong,
  and J.~Tang, ``Are we really making much progress? revisiting, benchmarking
  and refining heterogeneous graph neural networks,'' in \emph{Proceedings of
  the 27th ACM SIGKDD Conference on Knowledge Discovery \& Data Mining}, 2021,
  pp. 1150--1160.

\bibitem{kipf2017semi}
T.~N. Kipf and M.~Welling, ``Semi-supervised classification with graph
  convolutional networks,'' in \emph{ICLR}, 2017.

\bibitem{velivckovic2017graph}
P.~Veli{\v{c}}kovi{\'c}, G.~Cucurull, A.~Casanova, A.~Romero, P.~Lio, and
  Y.~Bengio, ``Graph attention networks,'' \emph{arXiv preprint
  arXiv:1710.10903}, 2017.

\bibitem{xu2018how}
K.~Xu, W.~Hu, J.~Leskovec, and S.~Jegelka, ``How powerful are graph neural
  networks?'' in \emph{ICLR}, 2019.

\bibitem{shirian2021dynamic}
A.~Shirian, S.~Tripathi, and T.~Guha, ``Dynamic emotion modeling with learnable
  graphs and graph inception network,'' \emph{IEEE Transactions on Multimedia},
  2021.

\bibitem{gemmeke2017audio}
J.~F. Gemmeke, D.~P. Ellis, D.~Freedman, A.~Jansen, W.~Lawrence, R.~C. Moore,
  M.~Plakal, and M.~Ritter, ``Audio set: An ontology and human-labeled dataset
  for audio events,'' in \emph{ICASSP}, 2017, pp. 776--780.

\bibitem{hershey2017cnn}
S.~Hershey, S.~Chaudhuri, D.~P. Ellis, J.~F. Gemmeke, A.~Jansen, R.~C. Moore,
  M.~Plakal, D.~Platt, R.~A. Saurous, B.~Seybold \emph{et~al.}, ``Cnn
  architectures for large-scale audio classification,'' in \emph{ICASSP}, 2017,
  pp. 131--135.

\bibitem{xie2017rethinking}
S.~Xie, C.~Sun, J.~Huang, Z.~Tu, and K.~Murphy, ``Rethinking spatiotemporal
  feature learning for video understanding,'' \emph{arXiv preprint
  arXiv:1712.04851}, vol.~1, no.~2, p.~5, 2017.

\bibitem{han2020self}
T.~Han, W.~Xie, and A.~Zisserman, ``Self-supervised co-training for video
  representation learning,'' \emph{NeurIPS}, vol.~33, pp. 5679--5690, 2020.

\bibitem{dai2017very}
W.~Dai, C.~Dai, S.~Qu, J.~Li, and S.~Das, ``Very deep convolutional neural
  networks for raw waveforms,'' in \emph{ICASSP}.\hskip 1em plus 0.5em minus
  0.4em\relax IEEE, 2017, pp. 421--425.

\bibitem{akbari2021vatt}
H.~Akbari, L.~Yuan, R.~Qian, W.-H. Chuang, S.-F. Chang, Y.~Cui, and B.~Gong,
  ``Vatt: Transformers for multimodal self-supervised learning from raw video,
  audio and text,'' \emph{arXiv preprint arXiv:2104.11178}, 2021.

\bibitem{shirian2022self}
A.~Shirian, K.~Somandepalli, and T.~Guha, ``Self-supervised graphs for audio
  representation learning with limited labeled data,'' \emph{arXiv preprint
  arXiv:2202.00097}, 2022.

\bibitem{kong2020panns}
Q.~Kong, Y.~Cao, T.~Iqbal, Y.~Wang, W.~Wang, and M.~D. Plumbley, ``Panns:
  Large-scale pretrained audio neural networks for audio pattern recognition,''
  \emph{IEEE/ACM Transactions on Audio, Speech, and Language Processing},
  vol.~28, pp. 2880--2894, 2020.

\bibitem{gong2021ast}
Y.~Gong, Y.-A. Chung, and J.~Glass, ``Ast: Audio spectrogram transformer,''
  \emph{arXiv preprint arXiv:2104.01778}, 2021.

\bibitem{simonyan2014very}
K.~Simonyan and A.~Zisserman, ``Very deep convolutional networks for
  large-scale image recognition,'' in \emph{{ICLR}}, Y.~Bengio and Y.~LeCun,
  Eds., 2015.

\bibitem{hong2020holmes}
S.~Hong, Y.~Xu, A.~Khare, S.~Priambada, K.~Maher, A.~Aljiffry, J.~Sun, and
  A.~Tumanov, ``Holmes: health online model ensemble serving for deep learning
  models in intensive care units,'' in \emph{ACM SIGKDD}, 2020, pp. 1614--1624.

\end{thebibliography}

\end{document}